\documentclass[12pt,titlepage]{article}
\usepackage{amsfonts}
\usepackage{amsmath}

\begin{document}

\title{On Duffin-Kemmer-Petiau particles with a mixed minimal-nonminimal
vector coupling and the nondegenerate bound states for the one-dimensional
inversely linear background}
\date{}
\author{A.S. de Castro\thanks{%
castro@pq.cnpq.br.} \\
\\
UNESP - Campus de Guaratinguet\'{a}\\
Departamento de F\'{\i}sica e Qu\'{\i}mica\\
12516-410 Guaratinguet\'{a} SP - Brazil}
\date{ }
\maketitle

\begin{abstract}
The problem of spin-0 and spin-1 bosons in the background of a general
mixing of minimal and nonminimal vector inversely linear potentials is
explored in a unified way in the context of the Duffin-Kemmer-Petiau theory.
It is shown that spin-0 and spin-1 bosons behave effectively in the same
way. An orthogonality criterion is set up and it is used to determine
uniquely the set of solutions as well as to show that even-parity solutions
do not exist.
\end{abstract}

\section{Introduction}

The first-order Duffin-Kemmer-Petiau (DKP) formalism \cite{pet}-\cite{kem}
describes spin-0 and spin-1 particles and has been used to analyze
relativistic interactions of spin-0 and spin-1 hadrons with nuclei as an
alternative to their conventional second-order Klein-Gordon and Proca
counterparts. The DKP formalism enjoys a richness of couplings not capable
of being expressed in the Klein-Gordon and Proca theories \cite{gue}-\cite%
{vij}. Although the formalisms are equivalent in the case of minimally
coupled vector interactions \cite{mr}-\cite{lun}, the DKP formalism opens
news horizons as far as it allows other kinds of couplings which are not
possible in the Klein-Gordon and Proca theories. Nonminimal vector
potentials, added by other kinds of Lorentz structures, have already been
used successfully in a phenomenological context for describing the
scattering of mesons by nuclei \cite{cla1}-\cite{cla2}. Nonminimal vector
coupling with a quadratic potential \cite{Ait}, with a linear potential \cite%
{kuli}, and mixed space and time components with a step potential \cite{ccc3}%
-\cite{ccc2} and a linear potential \cite{jpa} have been explored in the
literature. See also Ref. \cite{jpa} for a comprehensive list of references
on other sorts of couplings and functional forms for the potential
functions. In Ref. \cite{jpa} it was shown that when the space component of
the coupling is stronger than its time component the linear potential, a
sort of vector DKP oscillator, can be used as a model for confining bosons.

The problem of a particle subject to an inversely linear potential in one
spatial dimension ($\sim |x|^{-1}$), known as the one-dimensional hydrogen
atom, has received considerable attention in the literature (for a rather
comprehensive list of references, see \cite{xia}). This problem presents
some conundrums regarding the parities of the bound-state solutions. This
problem was also analyzed with the Klein-Gordon equation for a Lorentz
vector coupling \cite{spe}-\cite{bar1}. By using the technique of continuous
dimensionality the problem was approached with the Schr\"{o}dinger,
Klein-Gordon and Dirac equations \cite{mos}. In this last work it was
concluded that the Klein-Gordon equation, with the interacting potential
considered as a time component of a vector, provides unacceptable solutions
while the Dirac equation has no bounded solutions at all. On the other hand,
in a more recent work \cite{xia} the authors use connection conditions for
the eigenfunctions and their first derivatives across the singularity of the
potential, and conclude that only the odd-parity solutions of the Schr\"{o}%
dinger equation survive. The problem was also sketched for a Lorentz scalar
potential in the Dirac equation in \cite{ho} and \cite{asc6}, for a general
mixing of vector and scalar couplings in the Dirac equation \cite{asc7} and
in the Klein-Gordon equation \cite{asc8}, and for a pseudoscalar coupling in
the Dirac equation \cite{asc9}.

The main purpose of the present article is to report on the properties of
the DKP theory with time components of minimal and nonminimal vector
inversely linear potentials for spin-0 and spin-1 bosons in a unified way.
This sort of mixing, beyond its potential physical applications, shows to be
a powerful tool to obtain a deeper insight about the nature of the DKP
equation and its solutions as far as it explores the differences between
minimal and nonminimal couplings. The problem is mapped into an exactly
solvable Sturm-Liouville problem of a Schr\"{o}dinger-like equation. The
effective potential resulting from the mapping has the form of the Kratzer
potential \cite{kra} and the closed form solution for the bound states is
uniquely determined by requiring orthonormalizability. The results imply
that even-parity solutions to the one-dimensional DKP hydrogen atom do not
exist.

\section{Vector couplings in the DKP equation}

The DKP equation for a free boson is given by \cite{kem} (with units in
which $\hbar =c=1$)%
\begin{equation}
\left( i\beta ^{\mu }\partial _{\mu }-m\right) \psi =0  \label{dkp}
\end{equation}%
\noindent where the matrices $\beta ^{\mu }$\ satisfy the algebra $\beta
^{\mu }\beta ^{\nu }\beta ^{\lambda }+\beta ^{\lambda }\beta ^{\nu }\beta
^{\mu }=g^{\mu \nu }\beta ^{\lambda }+g^{\lambda \nu }\beta ^{\mu }$
\noindent and the metric tensor is $g^{\mu \nu }=\,$diag$\,(1,-1,-1,-1)$.
That algebra generates a set of 126 independent matrices whose irreducible
representations are a trivial representation, a five-dimensional
representation describing the spin-0 particles and a ten-dimensional
representation associated to spin-1 particles. The second-order Klein-Gordon
and Proca equations are obtained when one selects the spin-0 and spin-1
sectors of the DKP theory. A well-known conserved four-current is given by $%
J^{\mu }=\bar{\psi}\beta ^{\mu }\psi $\noindent $/2$ where the adjoint
spinor $\bar{\psi}$ is given by $\bar{\psi}=\psi ^{\dagger }\eta ^{0}$ with $%
\eta ^{0}=2\beta ^{0}\beta ^{0}-1$. The time component of this current is
not positive definite but it may be interpreted as a charge density. Then
the normalization condition $\int d\tau \,J^{0}=\pm 1$ can be expressed as%
\begin{equation}
\int d\tau \,\bar{\psi}\beta ^{0}\psi =\pm 2  \label{norm}
\end{equation}%
where the plus (minus) sign must be used for a positive (negative) charge.

With the introduction of vector interactions, the DKP equation can be
written as%
\begin{equation}
\left( i\beta ^{\mu }\partial _{\mu }-m-\beta ^{\mu }A_{\mu }^{\left(
1\right) }-i[P,\beta ^{\mu }]A_{\mu }^{\left( 2\right) }\right) \psi =0
\label{dkp2}
\end{equation}%
where $P$ is a projection operator ($P^{2}=P$ and $P^{\dagger }=P$) in such
a way that $\bar{\psi}[P,\beta ^{\mu }]\psi $ behaves like a vector under a
Lorentz transformation as does $\bar{\psi}\beta ^{\mu }\psi $. Once again $%
\partial _{\mu }J^{\mu }=0$ \cite{jpa}. Notice that the vector potential $%
A_{\mu }^{\left( 1\right) }$ is minimally coupled but not $A_{\mu }^{\left(
2\right) }$. If the terms in the potentials $A_{\mu }^{\left( 1\right) }$
and $A_{\mu }^{\left( 2\right) }$ are time-independent one can write $\psi (%
\vec{r},t)=\phi (\vec{r})\exp (-iEt)$, where $E$ is the energy of the boson,
in such a way that the time-independent DKP equation becomes%
\begin{equation}
\left[ \beta ^{0}\left( E-A_{0}^{\left( 1\right) }\right) +i\beta ^{i}\left(
\partial _{i}+iA_{i}^{\left( 1\right) }\right) -\left( m+i[P,\beta ^{\mu
}]A_{\mu }^{\left( 2\right) }\right) \right] \phi =0  \label{DKP10}
\end{equation}%
In this case \ $J^{\mu }=\bar{\phi}\beta ^{\mu }\phi /2$ does not depend on
time, so that the spinor $\phi $ describes a stationary state. Note that the
time-independent DKP equation is invariant under a simultaneous shift of $E$
and $A_{0}^{\left( 1\right) }$, such as in the Schr\"{o}dinger equation, but
the invariance does not maintain regarding $E$ and $A_{0}^{\left( 2\right) }$%
. It can be shown (see Ref. \cite{jpa}) that any two stationary states with
distinct energies and subject to the boundary conditions%
\begin{equation}
\int d\tau \,\partial _{i}\left( \bar{\phi}_{\kappa }\beta ^{i}\phi _{\kappa
^{\prime }}\right) =0  \label{Corto}
\end{equation}%
are orthogonal in the sense that $\int d\tau \,\bar{\phi}_{\kappa }\beta
^{0}\phi _{\kappa ^{\prime }}=0$, for $E_{\kappa }\neq E_{\kappa ^{\prime }}$%
. In addition, in view of (\ref{norm}) the spinors $\phi _{\kappa }$ and $%
\phi _{\kappa ^{\prime }}$ are said to be orthonormal if%
\begin{equation}
\int d\tau \,\bar{\phi}_{\kappa }\beta ^{0}\phi _{\kappa ^{\prime }}=\pm
2\delta _{E_{\kappa }E_{\kappa ^{\prime }}}  \label{orto8}
\end{equation}

The charge-conjugation operation changes the sign of the minimal interaction
potential,\ i.e.\textit{\ }changes the sign of \ $A_{\mu }^{\left( 1\right)
} $. This can be accomplished by the transformation $\psi \rightarrow \psi
_{c}=\mathcal{C}\psi =CK\psi $, where $K$ denotes the complex conjugation
and $C$ is a unitary matrix such that $C\beta ^{\mu }=-\beta ^{\mu }C$. The
matrix that satisfies this relation is $C=\exp \left( i\delta _{C}\right)
\eta ^{0}\eta ^{1}$. The phase factor $\exp \left( i\delta _{C}\right) $ is
equal to $\pm 1$, thus $E\rightarrow -E$. Note also that $J^{\mu
}\rightarrow -J^{\mu }$, as should be expected for a charge current.
Meanwhile $C$ anticommutes with $[P,\beta ^{\mu }]$ and the
charge-conjugation operation entails no change on $A_{\mu }^{\left( 2\right)
}$. The invariance of the nonminimal vector potential under charge
conjugation means that it does not couple to the charge of the boson. In
other words, $A_{\mu }^{\left( 2\right) }$ does not distinguish particles
from antiparticles. Hence, whether one considers spin-0 or spin-1 bosons,
this sort of interaction can not exhibit Klein's paradox.

For the case of spin 0, we use the representation for the $\beta ^{\mu }$\
matrices given by \cite{ned1}%
\begin{equation}
\beta ^{0}=%
\begin{pmatrix}
\theta & \overline{0} \\
\overline{0}^{T} & \mathbf{0}%
\end{pmatrix}%
,\quad \beta ^{i}=%
\begin{pmatrix}
\widetilde{0} & \rho _{i} \\
-\rho _{i}^{T} & \mathbf{0}%
\end{pmatrix}%
,\quad i=1,2,3  \label{rep}
\end{equation}%
\noindent where%
\begin{eqnarray}
\ \theta &=&%
\begin{pmatrix}
0 & 1 \\
1 & 0%
\end{pmatrix}%
,\quad \rho _{1}=%
\begin{pmatrix}
-1 & 0 & 0 \\
0 & 0 & 0%
\end{pmatrix}
\notag \\
&&  \label{rep2} \\
\rho _{2} &=&%
\begin{pmatrix}
0 & -1 & 0 \\
0 & 0 & 0%
\end{pmatrix}%
,\quad \rho _{3}=%
\begin{pmatrix}
0 & 0 & -1 \\
0 & 0 & 0%
\end{pmatrix}
\notag
\end{eqnarray}%
\noindent $\overline{0}$, $\widetilde{0}$ and $\mathbf{0}$ are 2$\times $3, 2%
$\times $2 \ and 3$\times $3 zero matrices, respectively, while the
superscript T designates matrix transposition. Here the projection operator
can be written as \cite{gue} $P=\left( \beta ^{\mu }\beta _{\mu }-1\right)
/3=\mathrm{diag}\,(1,0,0,0,0)$. In this case $P$ picks out the first
component of the DKP spinor. The five-component spinor can be written as $%
\psi ^{T}=\left( \psi _{1},...,\psi _{5}\right) $ in such a way that the
time-independent DKP equation for a boson constrained to move along the $x$%
-axis decomposes into%
\begin{equation*}
\left( \frac{d^{2}}{dx^{2}}+k^{2}\right) \phi _{1}=0
\end{equation*}%
\begin{equation}
\phi _{2}=\frac{1}{m}\left( E-A_{0}^{\left( 1\right) }+iA_{0}^{\left(
2\right) }\right) \,\phi _{1}  \label{dkp4}
\end{equation}%
\begin{equation*}
\phi _{3}=\frac{i}{m}\frac{d\phi _{1}}{dx},\quad \phi _{4}=\phi _{5}=0
\end{equation*}%
where%
\begin{equation}
k^{2}=\left( E-A_{0}^{\left( 1\right) }\right) ^{2}-m^{2}+\left(
A_{0}^{\left( 2\right) }\right) ^{2}  \label{k}
\end{equation}%
Meanwhile,
\begin{equation}
J^{0}=\frac{E-A_{0}^{\left( 1\right) }}{m}\,|\phi _{1}|^{2},\quad J^{1}=%
\frac{1}{m}\text{Im}\left( \phi _{1}^{\ast }\,\frac{d\phi _{1}}{dx}\right)
\label{corrente4}
\end{equation}%
It is worthwhile to note that $J^{0}$ becomes negative in regions of space
where $E<A_{0}^{\left( 1\right) }$ (a circumstance associated to Klein's
paradox) and that $A_{\mu }^{\left( 2\right) }$ does not intervene
explicitly in the current. With spinors satisfying (\ref{Corto}), i.e.%
\begin{equation}
\left. \left( \frac{d\phi _{1\kappa }^{\ast }}{dx}\phi _{1\kappa ^{\prime
}}-\phi _{1\kappa }^{\ast }\frac{d\phi _{1\kappa ^{\prime }}^{\ast }}{dx}%
\right) \right\vert _{x=x_{\inf }}^{x=x_{\sup }}=0  \label{Corto1}
\end{equation}%
where $\left[ x_{\inf },x_{\sup }\right] $ is the range of $x$, the
orthonormalization formula (\ref{orto8}) becomes%
\begin{equation}
\int\limits_{-\infty }^{+\infty }dx\,\,\frac{\frac{E_{\kappa }+E_{\kappa
^{\prime }}}{2}-A_{0}^{\left( 1\right) }}{m}\,\phi _{1\kappa }^{\ast }\phi
_{1\kappa ^{\prime }}=\pm \delta _{E_{\kappa }E_{\kappa ^{\prime }}}
\label{ORTO1}
\end{equation}%
regardless $A_{\mu }^{\left( 2\right) }$. Eq. (\ref{ORTO1}) is in agreement
with the orthonormalization formula for the Klein-Gordon theory in the
presence of a minimally coupled potential \cite{puk}. This is not
surprising, because, after all, both DKP equation and Klein-Gordon equation
are equivalent under minimal coupling.

For the case of spin 1, the $\beta ^{\mu }$\ matrices are \cite{ned2}%
\begin{equation}
\beta ^{0}=%
\begin{pmatrix}
0 & \overline{0} & \overline{0} & \overline{0} \\
\overline{0}^{T} & \mathbf{0} & \mathbf{I} & \mathbf{0} \\
\overline{0}^{T} & \mathbf{I} & \mathbf{0} & \mathbf{0} \\
\overline{0}^{T} & \mathbf{0} & \mathbf{0} & \mathbf{0}%
\end{pmatrix}%
,\quad \beta ^{i}=%
\begin{pmatrix}
0 & \overline{0} & e_{i} & \overline{0} \\
\overline{0}^{T} & \mathbf{0} & \mathbf{0} & -is_{i} \\
-e_{i}^{T} & \mathbf{0} & \mathbf{0} & \mathbf{0} \\
\overline{0}^{T} & -is_{i} & \mathbf{0} & \mathbf{0}%
\end{pmatrix}
\label{betaspin1}
\end{equation}%
\noindent where $s_{i}$ are the 3$\times $3 spin-1 matrices $\left(
s_{i}\right) _{jk}=-i\varepsilon _{ijk}$, $e_{i}$ are the 1$\times $3
matrices $\left( e_{i}\right) _{1j}=\delta _{ij}$ and $\overline{0}=%
\begin{pmatrix}
0 & 0 & 0%
\end{pmatrix}%
$, while\textbf{\ }$\mathbf{I}$ and $\mathbf{0}$\textbf{\ }designate the 3$%
\times $3 unit and zero matrices, respectively. In this representation $%
P=\,\beta ^{\mu }\beta _{\mu }-2=\mathrm{diag}\,(1,1,1,1,0,0,0,0,0,0)$, i.e.
$P$ projects out the four upper components of the DKP spinor. \noindent With
the spinor written as $\psi ^{T}=\left( \psi _{1},...,\psi _{10}\right) $,
and partitioned as%
\begin{equation*}
\psi _{I}^{\left( +\right) }=\left(
\begin{array}{c}
\psi _{3} \\
\psi _{4}%
\end{array}%
\right) ,\quad \psi _{I}^{\left( -\right) }=\psi _{5}
\end{equation*}%
\begin{equation}
\psi _{II}^{\left( +\right) }=\left(
\begin{array}{c}
\psi _{6} \\
\psi _{7}%
\end{array}%
\right) ,\quad \psi _{II}^{\left( -\right) }=\psi _{2}  \label{part}
\end{equation}%
\begin{equation*}
\psi _{III}^{\left( +\right) }=\left(
\begin{array}{c}
\psi _{10} \\
-\psi _{9}%
\end{array}%
\right) ,\quad \psi _{III}^{\left( -\right) }=\psi _{1}
\end{equation*}%
the one-dimensional time-independent DKP equation can be expressed as
\begin{equation*}
\left( \frac{d^{2}}{dx^{2}}+k^{2}\right) \phi _{I}^{\left( \sigma \right) }=0
\end{equation*}%
\begin{equation}
\phi _{II}^{\left( \sigma \right) }=\frac{1}{m}\left( E-A_{0}^{\left(
1\right) }+i\sigma A_{0}^{\left( 2\right) }\right) \,\phi _{I}^{\left(
\sigma \right) }  \label{spin1-ti}
\end{equation}%
\begin{equation*}
\phi _{III}^{\left( \sigma \right) }=\frac{i}{m}\frac{d\phi _{I}^{\left(
\sigma \right) }}{dx},\quad \phi _{8}=0
\end{equation*}%
where $k$ is again given by (\ref{k}), and $\sigma $ is equal to $+$ or $-$.
Now the components of the four-current are%
\begin{equation}
J^{0}=\frac{E-A_{0}^{\left( 1\right) }}{m}\sum\limits_{\sigma }|\phi
_{I}^{\left( \sigma \right) }|^{2},\quad J^{1}=\frac{1}{m}\text{Im}%
\sum\limits_{\sigma }\phi _{I}^{\left( \sigma \right) \dagger }\,\frac{d\phi
_{I}^{\left( \sigma \right) }}{dx}  \label{CUR2}
\end{equation}%
and the orthonormalization formula (\ref{orto8}) takes the form%
\begin{equation}
\int\limits_{-\infty }^{+\infty }dx\,\,\frac{\frac{E_{\kappa }+E_{\kappa
^{\prime }}}{2}-A_{0}^{\left( 1\right) }}{m}\,\sum\limits_{\sigma }\phi
_{I\kappa }^{\left( \sigma \right) \dagger }\phi _{I\kappa ^{\prime
}}^{\left( \sigma \right) }=\pm \delta _{E_{\kappa }E_{\kappa ^{\prime }}}
\end{equation}%
Just as for scalar bosons, $J^{0}<0$ for $E<A_{0}^{\left( 1\right) }$ and $%
A_{\mu }^{\left( 2\right) }$ does not appear in the current. Similarly, $%
A_{\mu }^{\left( 2\right) }$ do not manifest explicitly in the
orthonormalization formula. The prescribed orthonormalization expression is
well-founded provided%
\begin{equation}
\left. \sum\limits_{\sigma }\left( \frac{d\phi _{I\kappa }^{\left( \sigma
\right) \dagger }}{dx}\phi _{I\kappa ^{\prime }}^{\left( \sigma \right)
}-\phi _{I\kappa }^{\left( \sigma \right) \dagger }\frac{d\phi _{I\kappa
^{\prime }}^{\left( \sigma \right) }}{dx}\right) \right\vert _{x=x_{\inf
}}^{x=x_{\sup }}=0  \label{Corto2}
\end{equation}

Comparison between the two sets of formulas for the spin-0 and spin-1
sectors of the DKP theory evidences that vector bosons and scalar bosons
behave in a similar way.

\section{The inversely linear potential}

Now we are in a position to use the DKP equation with specific forms for
vector interactions. Let us focus our attention on time components of
minimal and nonminimal vector potentials in the inversely linear form, viz.
\begin{equation}
A_{0}^{\left( 1\right) }=-\frac{g_{1}}{|x|},\quad A_{0}^{\left( 2\right) }=-%
\frac{g_{2}}{|x|}  \label{sc1a}
\end{equation}%
where the coupling constants, $g_{1}$ and $g_{2}$, are real parameters. In
this case the first equations of (\ref{dkp4}) and (\ref{spin1-ti})
transmutes into%
\begin{equation}
-\frac{1}{2m}\frac{d^{2}\Phi }{dx^{2}}+V_{\mathtt{eff}}\,\Phi =E_{\mathtt{eff%
}}\,\Phi  \label{sc1b}
\end{equation}%
where $\Phi $ is equal to $\phi _{1}$ for the scalar sector, and to $\phi
_{I}^{\left( \sigma \right) }$ for the vector sector, with%
\begin{equation}
V_{\mathtt{eff}}=-\frac{q}{|x|}+\frac{\alpha }{x^{2}},\quad E_{\mathtt{eff}}=%
\frac{E^{2}-m^{2}}{2m}  \label{sc2}
\end{equation}%
and
\begin{equation}
q=\frac{E}{m}g_{1},\quad \alpha =-\frac{g_{1}^{2}+g_{2}^{2}}{2m}  \label{sc3}
\end{equation}

\noindent Therefore, one has to search for bounded solutions in an effective
Kratzer-like potential for $g_{1}\neq 0$, or in an inversely quadratic
potential for the case of a pure nonminimal vector potential ($g_{1}=0$).
Inasmuch as the origin is a singular point of (\ref{sc1b}) one could expect
the existence of singular solutions for $\Phi $. In all the circumstances
the effective potential presents a singularity at the origin given by $%
-1/x^{2}$. It is worthwhile to note at this point that the singularity $%
-1/x^{2}$ will never expose the particle to collapse to the center \cite{lan}
on the condition that $\alpha $ is greater than the critical value%
\begin{equation}
\alpha _{c}=-\frac{1}{8m}  \label{ac}
\end{equation}%
In the following this necessary condition for the existence of bound-state
solutions will be obtained in an alternative way. Note that the effective
potential could bind the particle only if $E_{\mathtt{eff}}<0$,
corresponding to energies in the range $|E|<m$.

The Schr\"{o}dinger equation with the Kratzer-like potential is an exactly
solvable problem and its solution, for an attractive inversely linear term
plus a repulsive inverse-square term in the potential, can be found on
textbooks \cite{lan}-\cite{flu}. Since we need solutions involving either a
repulsive or an attractive term in the inversely linear potential plus an
attractive inversely quadratic potential, the calculation including this
generalization is presented. Since $V_{\mathtt{eff}}$ is invariant under
reflection through the origin ($x\rightarrow -x$), eigenfunctions with
well-defined parities can be built. Thus, one can concentrate attention on
the positive half-line and impose boundary conditions on $\Phi $ at $x=0$
and $x=\infty $. Normalizability requires $\Phi \left( \infty \right) =0$
and the boundary condition at the origin will come into existence by
demanding orthogonality. As $x\rightarrow 0$, when the term $1/x^{2}$
dominates, the solution behaves as $x^{s}$, where $s$ is a solution of the
algebraic equation

\begin{equation}
s(s-1)-2m\alpha =0  \label{17}
\end{equation}%
viz.

\begin{equation}
s=\frac{1}{2}\left( 1\pm \sqrt{1+8m\alpha }\right)  \label{18}
\end{equation}

\noindent Due to the two-fold possibility of signs for $s$, it seems the the
solution of our problem can not be uniquely determined. However, the sine
qua non condition for orthogonality as dictated by (\ref{Corto1}) and (\ref%
{Corto2}) can be recast into a form similar to that one of the
nonrelativistic case \cite{xia}, \cite{cas}

\begin{equation}
\lim_{x\rightarrow 0}\left( \Phi _{\kappa }^{\ast }\frac{d\Phi _{\kappa
^{\prime }}}{dx}-\frac{d\Phi _{\kappa }^{\ast }}{dx}\Phi _{\kappa ^{\prime
}}\right) =0  \label{22-2}
\end{equation}%
and there results that the allowed values for $s$ are restricted to Re$%
\left( s\right) >1/2$ . Therefore, $\alpha >\alpha _{c}$ and the minus sign
in (\ref{18}) must be ruled out. That is to say that $s\ $is a real quantity
in the open interval with $1/2<s<1$, or equivalently $%
0<g_{1}^{2}+g_{2}^{2}<1/4$. Under those conditions the singular possibility
for $\Phi $ is kept away and $|\Phi |^{2}/|x|$ behaves better than $x^{-1}$
at the origin so that the square-integrability of $\Phi $, even if $g_{1}=0$%
, is ensured. This tells us that the behaviour of $\Phi $ at very small $x$
implies into the homogeneous Dirichlet condition $\Phi \left( 0\right) =0$.
We shall now distinguish the cases $g_{1}=0$ and $g_{1}\neq 0$.

\subsection{$g_{1}=0$}

Defining%
\begin{equation}
z=2\sqrt{-2mE_{\mathtt{eff}}}\,x  \label{z}
\end{equation}%
where the quantity under the radical sign is either positive or negative,
one obtains a special case of Whittaker%
\'{}%
s differential equation \cite{abr}%
\begin{equation}
\Phi ^{\prime \prime }+\left( -\frac{1}{4}-\frac{2m\alpha }{z^{2}}\right)
\Phi =0
\end{equation}%
The prime denotes differentiation with respect to $z$. The normalizable
asymptotic form of the solution as $z\rightarrow \infty $ is $e^{-z/2}$ with
$z>0$. Notice that this asymptotic behaviour rules out the possibility $E_{%
\mathtt{eff}}>0$, as has been pointed out already based on qualitative
arguments. The exact solution can now be written as%
\begin{equation}
\Phi =z^{s}w(z)e^{-z/2}  \label{comb1}
\end{equation}%
where $w$ is a regular solution of the confluent hypergeometric equation
(Kummer's equation) \cite{abr}
\begin{equation}
zw^{\prime \prime }+(b-z)w^{\prime }-aw=0  \label{35}
\end{equation}%
\noindent with the definitions%
\begin{equation}
a=s,\quad b=2s  \label{36}
\end{equation}%
The general solution of \ (\ref{35}) is expressed in terms of the confluent
hypergeometric functions (Kummer's functions) $_{1}F_{1}(a,b,z)$ (or $%
M(a,b,z)$) and $_{2}F_{0}(a,1+a-b,-1/z)$ (or $U(a,b,z)$):
\begin{equation}
w=A\,\,_{1}F_{1}(a,b,z)+Bz^{-a}\,_{2}F_{0}(a,1+a-b,-\frac{1}{z}),\quad b\neq
-\tilde{n}  \label{W}
\end{equation}%
where $\tilde{n}$ is a nonnegative integer. Due to the singularity of the
second term at $z=0$, only choosing $B=0$ gives a behavior at the origin
which can lead to square-integrable solutions. Furthermore, the requirement
of finiteness for $\Phi $ at $z=\infty $ implies that the remaining
confluent hypergeometric function ($_{1}F_{1}(a,b,z)$) should be a
polynomial. This is because $_{1}F_{1}(a,b,z)$ goes as $e^{z}$ as $z$ goes
to infinity unless the series breaks off. This demands that $a=-n$, where $n$
is also a nonnegative integer. This requirement combined with (\ref{36})
implies that the existence of bound-state solutions for pure inversely
quadratic potentials is out of question.

\subsection{$g_{1}\neq 0$}

As for $g_{1}\neq 0$, it is convenient to define the dimensionless quantity $%
\gamma $,

\negthinspace
\begin{equation}
\gamma =q\,\sqrt{-\frac{m}{2E_{\mathtt{eff}}}}  \label{15}
\end{equation}%
\noindent \noindent and using (\ref{sc1b})-(\ref{sc3}), with $z$ defined in (%
\ref{z}), one obtains the complete form for Whittaker%
\'{}%
s equation \cite{abr}

\begin{equation}
\,\Phi ^{\prime \prime }+\left( -\frac{1}{4}+\frac{\gamma }{z}-\frac{%
2m\alpha }{z^{2}}\right) \Phi =0  \label{16}
\end{equation}

\noindent Because $\alpha \neq 0$, the normalizable asymptotic form of the
solution as $z\rightarrow \infty $ is again given by $e^{-z/2}$ and $E_{%
\mathtt{eff}}<0$, i.e. $|E|<m$. The solution for all $z$ can be again
expressed as in (\ref{comb1}), but now $w$ is the regular solution of Kummer%
\'{}s equation \noindent with

\begin{equation}
a=s-\gamma ,\quad b=2s  \label{20}
\end{equation}

\noindent Then $w$ is expressed as $_{1}F_{1}(a,b,z)$ and in order to
furnish normalizable $\Phi $, the confluent hypergeometric function must be
a polynomial. This demands that $a=-n$, where $n$ is a nonnegative integer
in such a way that $_{1}F_{1}(a,b,z)$ is proportional to the associated
Laguerre polynomial $L_{n}^{\left( b-1\right) }(z)$, a polynomial of degree $%
n$. This requirement, combined with (\ref{20}), also implies into quantized
energies:

\begin{equation}
E=\,\varepsilon (g_{1})\,m\left\{ 1+\left[ \frac{g_{1}}{n+\frac{1}{2}+\sqrt{%
\frac{1}{4}-\left( g_{1}^{2}+g_{2}^{2}\right) }}\right] ^{2}\right\}
^{-1/2},\quad n=0,1,2,\ldots
\end{equation}

\noindent where $\varepsilon $, the sign function, is there because $%
Eg_{1}>0 $ due to the fact that $\gamma =n+s>0$ ($q>0$).

On the half-line, $\Phi $ is given by

\begin{equation}
\Phi (z)=Nz^{s}e_{\;}^{-z/2}\;L_{n}^{\left( 2s-1\right) }\left( z\right)
,\quad n=0,1,2,\ldots  \label{22}
\end{equation}

\noindent where $N$ is a constant related to the normalization.
Eigenfunctions on the whole line with well-defined parities can be built.
Those eigenfunctions can be constructed by taking symmetric and
antisymmetric linear combinations of $\Phi $. These new eigenfunctions
possess the same eigenenergy, then, in principle, there is a two-fold
degeneracy. Nevertheless, the matter is a little more complicated because
the effective potential presents a singularity. Since $\Phi $ vanishes at
the origin, the symmetric combination of $\Phi $ presents a discontinuous
first derivative at the origin. \noindent In fact, $\Phi $ is not
differentiable at the origin (recall that near the origin $\Phi $ behaves
like $x^{s}$ with $1/2<s<1$). In the specific case under consideration, the
effect of the singularity of the potential on $\Phi ^{\prime }=d\Phi /dx$
can be evaluated by integrating (\ref{sc1b}) from $-\delta $ to $+\delta $
and taking the limit $\delta \rightarrow 0$. The connection condition among $%
\Phi ^{\prime }(+\delta )$ and $\Phi ^{\prime }(-\delta )$ can be summarized
as

\begin{equation}
\Phi ^{\prime }(+\delta )-\Phi ^{\prime }(-\delta )=2m\int_{-\delta
}^{+\delta }dx\;V_{\mathtt{eff}}\,\Phi  \label{23f}
\end{equation}%
Substitution of (\ref{22}) into (\ref{23f}) allows us to conclude that only
the odd-parity combination furnishes a consistent result. This happens
because the right-hand side of (\ref{23f}) vanishes for an odd
eigenfunction, as it should do. For an even eigenfunction, though, the
right-hand side of (\ref{23f}) should equal $-2\Phi ^{\prime }(-\delta )$
for arbitrary $g_{1}$ and $g_{2}$, but it does not. Therefore, we are forced
to conclude that the $\Phi $ must be an odd-parity function. As an
unavoidable conclusion, the bound-state solutions are nondegenerate.

\section{Conclusions}

We succeed in searching for exact DKP bounded solutions for massive
particles by considering a mixing of minimal and nonminimal vector inversely
linear potentials for spin-0 and spin-1 bosons in a unified way. The
solution of the DKP-Coulomb problem was uniquely determined by requiring
orthonormalizability. As a bonus, the appropriate boundary conditions on $%
\Phi $ were proclaimed. A pure nonminimal coupling does not hold bound
states. For $g_{1}\neq 0$, there is an infinite set of bound-state solutions
either for particles (in the range $0<E<m$) or for antiparticles (in the
range $-m<E<0$). The spectrum does not distinguish the sign of $g_{2}$, but $%
E$ goes to $-E$ as $g_{1}\rightarrow -g_{1}$ as it has already been
anticipated by the charge-conjugate transformation arguments in Section 2.
No matter the signs of the potentials or how strong they are, the particle
and antiparticle levels neither meet nor dive into the continuum. Thus there
is no room for the production of particle-antiparticle pairs. This all means
that Klein\'{}s paradox never manifests. The regime of weak coupling ($%
0<g_{1}<<1/2$ and $|g_{2}|<<1/2$) runs in the nonrelativistic limit, viz. $%
E-m\simeq -mg_{1}^{2}/\left[ 2\left( n+1\right) ^{2}\right] $. This
nonrelativistic limit, where only the $g_{1}$-dependence survives,
corresponds to the energy levels for particles in a nonrelativistic
one-dimensional Coulomb potential \cite{xia}. Invariably, the spectrum is
nondegenerate and the eigenfunction behaves as an odd-parity function.

\bigskip \bigskip \bigskip \bigskip

\bigskip \bigskip \bigskip \bigskip

\bigskip

\noindent {\textbf{Acknowledgments}}

This work was supported in part by means of funds provided by Conselho
Nacional de Desenvolvimento Cient\'{\i}fico e Tecnol\'{o}gico (CNPq).

\end{document}